\documentclass[12pt]{article}
\usepackage[utf8x]{inputenc}
\usepackage{ucs}
\usepackage[russian,english,ukrainian]{babel}
\usepackage[OT1]{fontenc}
\usepackage{amsmath}
\usepackage{amsfonts}
\usepackage{amssymb}
\usepackage{graphicx}
\usepackage[numbers,sort&compress]{natbib}
\usepackage[bookmarks=true,colorlinks=true,citecolor=blue,linkcolor=blue,urlcolor=magenta]{hyperref}
\usepackage[left=1.5cm,right=1.5cm,top=2cm,bottom=2cm]{geometry}

\graphicspath{{noiseimages/}}
\usepackage{multirow}

\usepackage{xcolor}
\usepackage{soulutf8}
\definecolor{lightred}{rgb}{1,.90,.90}
\definecolor{pink}{rgb}{1,.80,.80}
\definecolor{lightgreen}{rgb}{0.9,1.0,0.9}
\definecolor{lightblue}{rgb}{.80,.90,1}
\definecolor{lightyellow}{rgb}{1.0,1.0,0.7}
\definecolor{lightorange}{rgb}{1.0,0.8,0.5}
\sethlcolor{lightyellow}
\sethlcolor{pink}

\begin{document}

\title{МОДЕЛЮВАННЯ ХВИЛЬОВОГО НЕЙТРОННО-ЯДЕРНОГО ГОРІННЯ $^{232}$Th ПРИ ЗБАГАЧЕННІ ПО $^{239}$Pu ДЛЯ ТЕПЛОВОЇ ОБЛАСТІ ЕНЕРГІЙ НЕЙТРОНІВ}

\author{А.О. Какаєв$^{1}$,  В.О. Тарасов$^{1}$, С.А. Чернеженко$^{1}$, В.О. Сова$^{2}$, В.Д. Русов$^{1}$}

\date{}


\maketitle
\begin{center}
\parbox{15cm}{
$^{1}$\textit{Одеський національний політехнічний університет,\\пр. Шевченко, 1, 65044, Одеса, Україна}

$^{2}$\textit{Державний науково-технічний центр з ядерної та радіаційної безпеки,\\вул. Василя Стуса 35-37, 03142, Київ, Україна}

\textit {e-mail:andreykakaev@gmail.com}}
\end{center}

\begin{abstract}
Розробка хвильових реакторів, які будуть працювати у режимі  хвильового ядерного горіння (ХЯГ), на початковій стадії, потребує дослідження кінетики режиму ХЯГ палива при зміні як зовнішніх параметрів (щільність потоку зовнішнього джерела нейтронів, теплофізичні параметри теплопереносу), так і внутрішніх параметрів (склад палива, матеріальний параметр реактора, запізнілі нейтрони).

Для підтвердження можливості ХЯГ при збагаченні, було проведено оцінку впливу на виконання критерію ХЯГ для торієвого палива (${}^{232}$Th при різних збагаченнях по ${}^{239}$Pu) і його поведінку на етапі підпалу. 

Для підтвердження справедливості виконання критерію повільного ХЯГ в залежності від енергії нейтронів, було 	проведено чисельне моделювання динаміки режиму ХЯГ торієвого палива з урахуванням запізнілих нейтронів в тепловій і надтеплових області енергій нейтронів (0.015-10~еВ). 
\end{abstract}

\textit{Ключові слова:} торієве реакторне паливо, критерій хвильового ядерного горіння, моделювання хвильового ядерного горіння, реактор, що працює у режимі хвильового ядерного горіння

\vspace{1cm}

\section{Вступ}

Робота присвячена дослідженню режимів хвильового ядерного горіння (ХЯГ) реакторного палива на основі торію-232 для  хвильових реакторів V покоління, які будуть працювати у режимі  хвильового ядерного горіння (ХЯГ). Таки ядерні реактори є реакторами з так званою внутрішньою безпекою. Основний принцип роботи реактора з внутрішньою безпекою виконується коли паливні компоненти структуровані так, щоб, по-перше, його характерний час регулювання був набагато більше хвилини, по-друге, щоб в режимі його роботи з'явилися елементи саморегулювання \cite{1,2,3,4,5}. 

Цього можна досягти, якщо в активній зоні реактора, крім інших реакцій, має місце наступний ланцюжок перетворень:

\begin{equation} 
{}^{232}Th(n,\gamma )\to {}^{233}Th\stackrel{\beta ^{-} }{\longrightarrow} {}^{233}Pa\stackrel{\beta ^{-} }{\longrightarrow} {}^{233}U .  
\label{eq01}
\end{equation} 

\noindent
де ${}^{232}$Th, ${}^{233}$U, ${}^{233}$Pa є відповідні ізотопи торію, урану і протактинію.

У цьому випадку утворюється ${}^{233}$U, який є основним компонентом подільної речовини. Характерний час цієї реакції відповідає часу двох $\beta^{-}$ розпадів, що дорівнює приблизно $\tau = 27.4 / \ln 2 = 39.53$~діб, що на кілька порядків більше, ніж для запізнілих нейтронів, що в свою чергу дає нам чимало часу для маневрування реактором. Дане перетворення і буде розглянуто в рамках даної роботи.

Ефект саморегулювання пов'язаний з тим, що збільшення потоку нейтронів призведе до швидкого вигоряння ${}^{233}$U, зменшення його концентрації і відповідно потоку нейтронів (утворення нових ядер ${}^{233}$U буде йти в колишньому темпі приблизно протягом 40 діб). Якщо ж, навпаки, потік нейтронів в результаті зовнішнього втручання зменшиться, то знижується швидкість вигоряння і збільшиться темп напрацювання подільної речовини (в даному випадку ${}^{233}$U) з подальшим збільшенням числа виділених нейтронів в реакторі через приблизно такий самий час.

Досить повна математична модель активної зони реактора повинна включати в себе моделі нестаціонарних тривимірних процесів переносу нейтронів в сильно неоднорідному середовищі, вигоряння палива і реакторної кінетики, а також модель відведення тепла.

В роботах \cite{1,2} для U-Pu паливного середовища було розглянуто можливість реалізації режиму ХЯГ при деяких спрощеннях кінетичної системи рівнянь, що описують режим ХЯГ: розглядається одновимірне середовище, фіксована енергія нейтронів (одногрупове наближення); не враховується дифузія нейтронів; кінетичне рівняння для ${}^{239}$Pu написано в припущенні, що ${}^{238}$U безпосередньо переходить в ${}^{239}$Pu з деяким характерним часом $\beta $ розпаду $\tau _{\beta } $; не враховуються запізнілі нейтрони і температура середовища, що поділяється. 

У роботі \cite{1} отримані вирази для нерівноважно-стаціонарної $N_{N{\rm .}S{\rm .}}^{Pu} $ (в \cite{1} рівноважної) і критичної $N_{crit}^{Pu}$ концентрації нукліда (${}_{{\rm 94}}^{{\rm 239}} Pu$), що поділяється. Ці вирази були адаптовані для задачі, що розглядається у даній роботі, а саме, було замінено нуклід ${}^{239}$Pu, що поділяється, на нуклід ${}^{233}$U. Відповідно, перше завантаження палива складається з ${}^{232}$Th з невеликим збагаченням по ${}^{239}$Pu, тому згідно з відповідним ланцюжком розпаду (\ref{eq01}), отримуємо:

\begin{equation} 
N_{N{\rm .}S{\rm .}}^{U{\rm 233}} \left(E_{n} \right)\approx \frac{\sigma _{c}^{Th{\rm 232}} \left(E_{n} \right)}{\sigma _{c}^{U{\rm 233}} \left(E_{n} \right)+\sigma _{f}^{U233} \left(E_{n} \right)} N^{Th{\rm 232}} 
=\frac{\sigma _{c}^{Th{\rm 232}} \left(E_{n} \right)}{\sigma _{a}^{U{\rm 233}} \left(E_{n} \right)} N^{Th{\rm 232}}  
\label{eq02}
\end{equation} 

\begin{equation}
N_{crit}^{U{\rm 233}} \left(E_{n} \right)\approx \frac{\sum _{i\ne U{\rm 233}}\sigma _{a}^{i} \left(E_{n} \right) N^{i} -\sum _{i\ne U{\rm 233}}\nu _{i} \sigma _{f}^{i} \left(E_{n} \right)N^{i}  }{(\nu _{U{\rm 233}} -{\rm 1})\sigma _{f}^{U{\rm 233}} \left(E_{n} \right)-\sigma _{c}^{U{\rm 233}} \left(E_{n} \right)}  
\label{eq03}
\end{equation} 

\noindent
де $\sigma _{c}^{i} ,\sigma _{f}^{i} ,\sigma _{a}^{i}$ - мікроперерізи реакцій радіаційного захоплення нейтрона, поділу і поглинання, відповідно для \textit{i }--го нукліду середовища, що поділяється; $\tau _{\beta }$- характерний час для двох $\beta$- розпадів, що перетворюють ${}_{90}^{233}$Th (${}_{90}^{233}$Th утворюється при радіаційному захопленні нейтронів ${}_{90}^{232}$Th) в ${}_{91}^{233}$Pa, а останній у ${}_{91}^{233}$U; $\nu _{i} $ і $\nu _{U233} $- середнє число нейтронів, що народжуються при поділі одного ядра \textit{i}-го нукліда і ${}_{91}^{233}$U відповідно.

Критерій ХЯГ має вигляд умови $N_{N{\rm .}S{\rm .}}^{U{\rm 233}} $$>$$N_{crit}^{U{\rm 233}} $ та його вперше було сформульовано у \cite{1}. Наочно розвиток подій можна уявити собі таким чином. Нейтрони, що випускаються зовнішнім джерелом, у паливному середовищі на відстані довжини пробігу поглинаються ${}^{232}$Th і утворюють ${}^{233}$U, що поділяється. По мірі накопичення ${}^{233}$U у найближчій до зовнішнього джерела нейтронів області палива процеси поділу ядер ${}^{233}$U посилюються, та при виконанні умови $N_{N{\rm .}S{\rm .}}^{U{\rm 233}} $$>$$N_{crit}^{U{\rm 233}} $ внаслідок швидкого розвитку ланцюгового процесу, у ній реалізується режим ядерного горіння. В свою чергу внаслідок реалізації режиму ядерного горіння у локальній області палива, нейтронів у ній стає досить, щоб запалити сусідню до неї, але більш віддалену від зовнішнього джерела нейтронів область палива. Після того, як центр енерговиділення зміщується вглиб, послаблюється роль зовнішнього джерела, система поступово виходить в стаціонарний режим.
\par Амереканська компанія ``Terra-power'' перша заявила про розробку швидкого ядерного реактора, що працює в режимі ХЯГ. Цей реактор отримав назву TWR (Traveling-Wave Reactor) \cite{6,7,8,9,10}. Навіть були оформлені патенти на реактори TWR \cite{10,11,12,13,14}, засновані на ідеї хвилі ядерного горіння, що біжить по паливу \cite{1,2,3,4,5,6,7,8,9}. Але в ході наукових обговорень про реалізацію проекту реактора TWR було виявлено деякі проблеми технічного характеру, головна з яких -- радіаційна стійкість конструкційного матеріалу першої стінки ТВЕЛа, тому найбільш важливим для технічної реалізації хвильових реакторів представляється необхідність пошуку рішення проблеми дії високої інтегральної дози швидких нейтронів на конструкційні матеріали, що іх пошкоджує, у хвильових ядерних реакторів. Як свідчать результати математичних моделювань (наприклад, \cite{3}) для режимів хвильового ядерного горіння радіаційна дія на конструкційні матеріали може досягати $\sim$500 ЗНА (зміщень на атом). Причому, на сьогодні конструкційних матеріалів, які витримують такий радіаційний вплив, поки ще не створено, і максимально досягнуте радіаційне навантаження для діючих реакторних металів дорівнює 100 ЗНА. Вирішенням даної проблеми є реалізація режиму XЯГ не на швидких нейтронах, а на теплових нейтронах або надтеплових нейтронах \cite{15}.


\section{Вплив збагачення по ${}^{239}$Pu на виконання критерію ХЯГ в торій-урановому подільному середовищі при енергіях нейтронів 0.015-10~еВ}
\label{sec01}

\begin{figure}[ht]
\center{\includegraphics[width=1\linewidth]{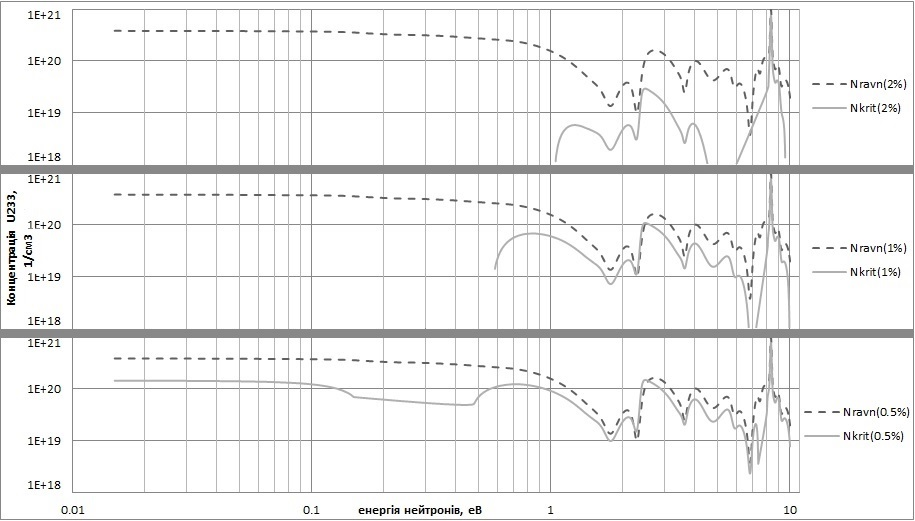}}
\vskip-3mm\caption{Залежність нерівноважно-стаціонарної та критичної концентрації ${}^{233}$U по енергії нейтронів в шкалі від 0.015$\div$10~еВ, який першочергово складається з ${}^{232}$Th зі збагаченням: 
a) 2\%; b) -- 1\%; c) --0.5\% по ${}^{239}$Pu.}
\label{fig01}
\end{figure}

Для паливного середовища із ${}^{232}$Th, що збагачено  ${}^{239}$Pu,  та для області енергій нейтронів 0.015-10~еВ відповідно до рівнянь (\ref{eq02}) і (\ref{eq03}) були проведені розрахунки нерівноважно-стаціонарної і критичної концентрацій ${}^{233}$U. Результати розрахунків представлені на рис.~\ref{fig01}.

Важливим моментом є первинне бомбардування палива нейтронами, тобто миттєвий поділ ядер ${}^{239}$Pu нейтронами, для насичення активної зони необхідним потоком нейтронів необхідним для ХЯГ. Такий процес зменшить час використання зовнішнього джерела нейтронів, який застосовують при первинному запуску будь-якого реактора, а також надасть можливості підтримання ядерної реакції до виходу її у стаціонарний режим. 

Результати, представлені на рис.~\ref{fig01}, свідчать про те, що для паливного середовища, яке складається з ${}^{232}$Th із збагаченням 2\% по ${}^{239}$Pu (рис.~\ref{fig01}а), критична концентрація ${}^{233}$U більше нуля лише для інтервалу енергій нейтронів $1\div$10~еВ. Для інших діапазонів енергій нейтронів критерій ХЯГ не виконується. Таким чином, якщо сформувати такий склад палива, щоб спектр нейтронів підтримувався постійно в зазначеному інтервалі енергій, то можлива реалізація такого хвильового ядерного реактора. Але в подальшому для реалізації стаціонарного ХЯГ потрібно підбирати склад та структуру активної зони реактора, яка повинна містити сповільнювач нейтронів.

Аналогічний аналіз представлених на (рис.~\ref{fig01}b,c) результатів дозволяє зробити висновок про те, що при зменшенні збагачення від 1\% до 0.5\% по ${}^{239}$Pu, відбувається розширення області енергій нейтронів, в якій критична концентрація ${}^{233}$U більше нуля, в сторону теплових енергій нейтронів, причому при концентрації 0.5\% по ${}^{239}$Pu, у всіх цих областях енергій нейтронів виконується критерій ХЯГ, тобто, можлива реалізація режиму хвильового нейтронно-ядерного горіння на теплових нейтронах.

Оскільки зі зменшенням збагачення область енергій нейтронів, в яких виконуються критерій ХЯГ, розширюється, відповідно можна зробити припущення, що при збагаченні ${}^{232}$Th по ${}^{239}$Pu слід виконувати тільки для «підпалу» ланцюгової реакції, або кажучи іншими словами, для насичення активної зони необхідним потоком нейтронів. При опроміненні нейтронами нейтронного джерела паливного середовища буде відбуватися поділ ${}^{239}$Pu, і по мірі його вигоряння, це також дає нейтрони для активації ${}^{232}$Th, який після захвату нейтронів переходить у ${}^{233}$Th, який через 39.5~діб перейде в потрібний нам нуклід ${}^{233}$U, що поділяється. Даний ізотоп урану в свою чергу почне поділ під дією нейтронів. Після поступового вигоряння ${}^{239}$Pu до концентрації менше 0.5\%  від загального завантаження палива (рис.~\ref{fig01}) і насичення активної зони потрібним потоком нейтронів, ланцюгова ядерна реакція зможе підтримувати сама себе, і відповідно, реалізується стаціонарна ХЯГ.


\section{Моделювання нейтронно-ядерного горіння торію-232 для теплової області енергії нейтронів}
\label{sec02}

Для підтвердження справедливості вищенаведених оцінок і висновків, заснованих на аналізі виконання критерію повільного хвильового нейтронно-ядерного горіння в залежності від енергії нейтронів, було проведено чисельне моделювання нейтронно-ядерного горіння ${}^{232}$Th в тепловій області енергій нейтронів (0.015-10 еВ).

Розглянемо напівпростір по координаті \textit{x}, заповнений ${}^{232}$Th (99.5\% ${}^{232}$Th і 0.5\% ${}^{239}$Pu), котрий опромінюється з відкритої поверхні нейтронним джерелом. Торій-232, якщо він поглинає нейтрон, перетворюється на ${}^{233}$Th, який потім внаслідок двох $\beta $-розпадів з характерним часом $\tau _{\beta } $- 39.5~діб переходить в ізотоп ${}^{233}$U. Як показано вище, в такому середовищі може виникнути повільна хвиля нейтронно-подільного горіння.

З урахуванням запізнілих нейтронів кінетика такої хвилі описується системою із 19 диференційних рівнянь у вигляді частинних похідних зі зворотними нелінійними зв'язками щодо 19 функцій $n\;\left(x,t\right)$, $N_{Pu} \left(x,t\right)$, $N_{Th{\rm 2}} \left(x,t\right)$, $N_{Th{\rm 3}} \left(x,t\right)$, $N_{U3} \left(x,t\right)$, $\tilde{N}_{i}^{\left(U3\right)} \left(x,t\right)$, $\tilde{N}_{i}^{\left(Pu\right)} \left(x,t\right)$, $\bar{N}^{\left(U3\right)} \left(x,t\right)$, $\bar{N}^{\left(Pu\right)} \left(x,t\right)$ двох змінних \textit{x} і \textit{t}, яка може бути записана наступним чином.

Спочатку випишемо кінетичне рівняння для щільності потоку нейтронів:
\begin{equation}
\frac{\partial \; n\; \left(x,t\right)}{\partial \; t} =D\; \Delta \; n\; \left(x,t\right)+q\; \left(x,t\right) 
\end{equation} 

\noindent
де об'ємна щільність джерела:

\begin{align}
q(x,t) & = \Bigg[\nu^{(U3)} ({\rm 1}-p^{(U3)} )-{\rm 1}]\cdot n(x,t)\cdot V_{n} \cdot \sigma_{f}^{U3} \cdot N_{U3} (x,t) \nonumber \\
& +[\nu ^{(Pu)} ({\rm 1}-p^{(Pu)} )-{\rm 1}]\cdot n(x,t)\cdot V_{n} \sigma _{f}^{Pu} \cdot N_{Pu} (x,t)+  \nonumber \\
& +{\rm ln2}\cdot \sum _{i={\rm 1}}^{{\rm 6}}[\frac{\tilde{N}_{i^{(U3)} } }{T_{{\rm 1}/{\rm 2}}^{i^{(U3)} } } +\frac{\tilde{N}_{i^{(Pu)} } }{T_{{\rm 1}/{\rm 2}}^{i^{(Pu)} } }] - n(x,t)\cdot V_{n} \cdot  \nonumber \\
& \cdot \bigg[\sum _{Pu,Th{\rm 2,}Th{\rm 3,}U3}\sigma _{c}^{i} \cdot N_{i} (x,t)+\sum _{i={\rm 1}}^{{\rm 6}}[\sigma _{c}^{i\; (U3)} \cdot \tilde{N}_{i}^{(U3)}(x,t)+\sigma _{c}^{i\; (U5)} \cdot \tilde{N}_{i_{i} }^{Pu} (x,t)] \bigg] \Bigg]  \nonumber \\
& -n(x,t)\cdot V_{n} \cdot \left[\sigma _{c}^{eff\left(U3\right)} \cdot \bar{N}^{\left(U3\right)} (x,t)+\sigma _{c}^{eff\left(Pu\right)} \cdot \bar{N}^{\left(Pu\right)} (x,t)\right] 
\label{eq05}
\end{align} 

\noindent
де $n\; \left(x,t\right)$ -- щільність нейтронів; $D$ -- коефіцієнт дифузії нейтронів; \textit{V${}_{n}$} -- швидкість нейтронів; $\nu ^{\left(U3\right)} $ і $\nu ^{\left(Pu\right)} $ дорівнюють середньому числу миттєвих нейтронів на один акт поділу ${}^{233}$U і ${}^{239}$Pu відповідно; $N_{Pu} $,$N_{Th{\rm 2}} $,$N_{Th{\rm 3}} $,$N_{U3} $ -- концентрації ${}^{239}$Pu, ${}^{232}$Th, ${}^{233}$Th, ${}^{233}$U відповідно; $\tilde{N}_{i}^{\left(U3\right)} $і $\tilde{N}_{i}^{\left(Pu\right)} $ -- концентрації нейтронно-надлишкових уламків поділу ядер ${}^{233}$U і ${}^{239}$Pu; $\bar{N}_{i}^{\left(U3\right)} $ і $\bar{N}_{i}^{\left(Pu\right)} $ -- концентрації всіх інших уламків поділу ядер ${}^{233}$U і ${}^{239}$Pu; $\sigma _{c}$ і $\sigma _{f}$ -- мікроперерізи реакцій радіаційного захоплення нейтрона та поділу ядра; параметри $p_{i} (p= \sum _{i=1}^{6} p_i)$ і $T_{1/2}^{i}$, характеризують групи запізнілих нейтронів для основних паливних нуклідів, взяті з \cite{16,17,18}. Відзначимо, що при виводі рівняння для $q(x,t)$ для врахування запізнілих нейтронів використовувався метод Ахієзера-Померанчука \cite{18}.

Останні члени в квадратних дужках в правій частині (\ref{eq05}) задавалися відповідно до методу усередненого ефективного перерізу для шлаків \cite{17}:  

\begin{equation}
n(x,t)V_{n} \sum _{i=fission\; fragments}\sigma _{c}^{i} \bar{N}_{i} (x,t)=n(x,t)V_{n} \sigma _{c}^{eff} \bar{N}(x,t) 
\end{equation} 

\noindent
де $\sigma _{c}^{eff} $ -- деякий ефективний мікропереріз радіаційного захоплення нейтронів для уламків.

Кінетичні рівняння для$\bar{N}^{\left(U3\right)} $(\textit{x}, \textit{t}) і $\bar{N}^{\left(Pu\right)} $(\textit{x}, \textit{t}) мали наступну форму:
\begin{equation}
\frac{\partial \bar{N}^{\left(U3\right)} (x,t)}{\partial t} ={\rm 2}\left({\rm 1}-\sum _{i={\rm 1}}^{{\rm 6}}p_{i}^{\left(U3\right)}  \right)\cdot n(x,t)\cdot V_{n} \cdot \sigma _{f}^{U3} \cdot N_{U3} (x,t)+\sum _{i={\rm 1}}^{{\rm 6}}\frac{\tilde{N}_{i}^{(U3)} {\rm ln2}}{T_{{\rm 1}/{\rm 2}}^{i\; (U3)} }   
\end{equation} 

і

\begin{equation}
\frac{\partial \bar{N}^{\left(Pu\right)} (x,t)}{\partial t} ={\rm 2}\left({\rm 1}-\sum _{i={\rm 1}}^{{\rm 6}}p_{i}^{\left(Pu\right)}  \right)\cdot n(x,t)\cdot V_{n} \cdot \sigma _{f}^{Pu}  \cdot N_{U5} (x,t)+\sum _{i={\rm 1}}^{{\rm 6}}\frac{\tilde{N}_{i}^{(Pu)} {\rm ln2}}{T_{{\rm 1}/{\rm 2}}^{i\; (Pu)} }   
\end{equation} 

Отже, маємо таку систему з 19 кінетичних рівнянь:
\begin{equation}
\frac{\partial \; n\; \left(x,t\right)}{\partial \; t} =D\; \Delta \; n\; \left(x,t\right)+q\left(x,t\right) 
\label{eq09}
\end{equation} 

\noindent
де $q(x,t)$ задається виразом (\ref{eq05});

\begin{equation}
\frac{\partial \; N_{Th{\rm 2}} \left(x,t\right)}{\partial \; t} =-\; V_{n} n\left(x,t\right)\; \sigma _{c}^{Th{\rm 2}} \; N_{Th{\rm 2}} \left(x,t\right); 
\end{equation} 

\begin{equation}
\frac{\partial \; N_{Th{\rm 3}} \left(x,t\right)}{\partial \; t} =V_{n} n\left(x,t\right)\; \left [\; \sigma _{c}^{Th{\rm 2}} \; N_{Th{\rm 2}} \left(x,t\right)-\sigma_{c}^{Th{\rm 3}} \; N_{Th{\rm 3}} \left(x,t\right)\right]\;-\frac{{\rm 1}}{\tau _{\beta } } N_{Th{\rm 3}} (x,t);
\end{equation} 

\begin{equation}
\frac{\partial \; N_{U3} \left(x,t\right)}{\partial \; t} =\frac{{\rm 1}}{\tau _{\beta } } N_{Th{\rm 3}} \left(x,t\right)-\; V_{n} n\left(x,t\right)\left(\sigma _{f}^{U3} +\sigma _{c}^{U3} \right)\; N_{U3} \left(x,t\right)\; ; 
\end{equation} 

\begin{equation}
\frac{\partial \; N_{Pu} \left(x,t\right)}{\partial \; t} =-\; V_{n} n\left(x,t\right)\; \left(\sigma _{f}^{Pu} +\sigma _{c}^{Pu} \right)\; N_{Pu} \left(x,t\right); 
\end{equation} 

\begin{equation} 
\frac{\partial \tilde{N}_{i}^{\left(U3\right)} \left(x,t\right)}{\partial t} =p_{i}^{\left(U3\right)} \cdot V_{n} \cdot n\; \left(x,t\right)\; \cdot \sigma _{f}^{U3} \cdot N_{U3} \left(x,t\right)-\frac{{\rm ln2}\cdot \tilde{N}_{i}^{\left(U3\right)} \left(x,t\right)}{T_{{\rm 12}}^{i\; \left(U3\right)} } ,\quad i={\rm 1,}\; {\rm 6} 
\end{equation} 

\begin{equation}
\frac{\partial \tilde{N}_{i}^{\left(Pu\right)} \left(x,t\right)}{\partial t} =p_{i}^{\left(Pu\right)} \cdot V_{n} \cdot n\; \left(x,t\right)\; \cdot \sigma _{f}^{Pu} \cdot N_{Pu} \left(x,t\right)-\frac{{\rm ln2}\cdot \tilde{N}_{i}^{\left(Pu\right)} \left(x,t\right)}{T_{{\rm 12}}^{i\; \left(Pu\right)} } ,\quad i={\rm 1,}\; {\rm 6} 
\end{equation} 

\begin{equation}
\frac{\partial \bar{N}^{\left(U3\right)} (x,t)}{\partial t} ={\rm 2}\left({\rm 1}-\sum _{i={\rm 1}}^{{\rm 6}}p_{i}^{\left(U3\right)}  \right)\cdot n(x,t)\cdot V_{n} \cdot \sigma _{f}^{U3}\cdot N_{U3} (x,t)+\sum _{i={\rm 1}}^{{\rm 6}}\frac{\tilde{N}_{i}^{\left(U3\right)} {\rm ln2}}{T_{{\rm 1}/{\rm 2}}^{i\; \left(U3\right)} }   
\end{equation} 

\begin{equation}
\frac{\partial \bar{N}^{\left(Pu\right)} (x,t)}{\partial t} ={\rm 2}\left({\rm 1}-\sum _{i={\rm 1}}^{{\rm 6}}p_{i}^{\left(Pu\right)}  \right)\cdot n(x,t)\cdot V_{n} \cdot \sigma _{f}^{Pu} \cdot N_{Pu} (x,t)+\sum _{i={\rm 1}}^{{\rm 6}}\frac{\tilde{N}_{i}^{\left(Pu\right)} {\rm ln2}}{T_{{\rm 1}/{\rm 2}}^{i\; \left(Pu\right)} }   
\label{eq17}
\end{equation} 

\noindent
де $\tau _{\beta} $ -- час життя ядра відносно $\beta$-розпаду.

Граничні умови:
 
\begin{equation}
n\;\left(x,t\right)|_{x={\rm 0}} =\frac{\Phi _{{\rm 0}} }{V_{n} }; \qquad n\;\left(x,t\right)|_{x=l} ={\rm 0}
\label{eq18}
\end{equation} 

\noindent
де $\Phi_0 $ -- щільність нейтронів, що створюються пласким дифузійним джерелом нейтронів, розташованим на границі при $x=0$; $l$ -- довжина блоку з природного урану, що задається при моделюванні.

Початкові умови:
\begin{equation}
 n\; \left(x,t\right)|_{x={\rm 0,}t={\rm 0}} =\frac{\Phi _{{\rm 0}} }{V_{n} }; \qquad n\; \left(x,t\right)|_{x\ne {\rm 0,}\; t={\rm 0}} ={\rm 0}
\end{equation} 

\begin{multline} 
N_{Th{\rm 2}} \left(x,t\right)|_{t={\rm 0}} ={\rm 0,995}\cdot \frac{\rho _{Th{\rm 2}} }{\mu _{Th{\rm 2}} } N_{A} \approx {\rm 0,995}\cdot \frac{{\rm 11},{\rm 78}}{{\rm 232}} N_{A};\\
 \quad N_{Pu} \left(x,t\right)|_{t={\rm 0}} \approx {\rm 0,005}\cdot \frac{{\rm 19},{\rm 8}4}{{\rm 239}} N_{A}
\end{multline} 

\noindent
де $\rho _{Th{\rm 2}} $ -- густина (г/см${}^{3}$) торію-232, $\mu _{Th{\rm 2}} $ -- моль (г/моль${}^{-1}$) торію-232, $N_{A} $ - число Авогадро;

\begin{multline}
{N}_{Th{\rm 3}} \left(x,t\right)|_{t={\rm 0}} ={\rm 0,}\quad {N}_{U3} (x,t)|_{t={\rm 0}} ={\rm 0,}\\
\tilde{N}_{i}^{\left(U3\right)} (x,t)|_{t={\rm 0}} ={\rm 0,}\quad \tilde{N}_{i}^{\left(Pu\right)} (x,t)|_{t={\rm 0}} ={\rm 0}\;\\
 \bar{N}_{i}^{\left(U3\right)} (x,t)|_{t={\rm 0}} ={\rm 0,}\quad \bar{N}_{i}^{\left(Pu\right)} (x,t)|_{t={\rm 0}} ={\rm 0} 
\label{eq21}
\end{multline} 

Чисельне рішення системи рівнянь (\ref{eq09}) - (\ref{eq17}) з граничними і початковими умовами (\ref{eq18}) - (\ref{eq21}) проводилося за допомогою програмного пакета Mathematica 8.

Для оптимізації процесу чисельного рішення системи рівнянь, було здійснено перехід до безрозмірних величин, згідно з наступними співвідношеннями:

\begin{equation} 
n\; \left(x,t\right)=\frac{\Phi _{{\rm 0}} }{V_{n} } n(x,t), N(x,t)=\frac{\rho _{Th{\rm 2}} N_{A} }{\mu _{Th{\rm 2}} } N(x,t)
\label{eq22}
\end{equation}

\begin{table}[ht!]
\noindent
\caption{Значення постійних коефіцієнтів диференційних рівнянь \cite{15,16,17,18}}
\begin{center}
\begin{tabular}{|p{2in}|p{0.9in}|p{0.7in}|p{0.7in}|p{0.7in}|} \hline 
\multicolumn{2}{|p{1in}|}{Характеристики нуклідів} & ${}^{232}$Th--${}^{233}$U & \multicolumn{2}{|p{1.2in}|}{99.5\% (${}^{232}$Th -- ${}^{233}$U) + 0.5\% ${}^{239}$Pu} \\ \hline 
Дифузія нейтрона & D, cм${}^{2}$/c & $2 \cdot 10^{-4}$ & \multicolumn{2}{|p{1.2in}|}{$2 \cdot 10^{-4}$} \\ \hline 
Швидкість нейтрона & V${}_{n}$, cм/c & $10^{6}$ & \multicolumn{2}{|p{1.2in}|}{$10^{6}$} \\ \hline 
Потік нейтронів & $\Phi_0$, 1/(cм${}^{2}$·c) & $10^{16}$ & \multicolumn{2}{|p{1.2in}|}{$10^{16}$} \\ \hline 
Переріз захоплення ${}^{232}$Th & $\sigma_c$ , барн & 3.06 & \multicolumn{2}{|p{1.2in}|}{3.06} \\ \hline 
Переріз захоплення ${}^{233}$Th & $\sigma_c$, барн & 14.59 & \multicolumn{2}{|p{1.2in}|}{14.59} \\ \hline 
Час двох бета розпадів & $\tau $, діб & 39.5 & \multicolumn{2}{|p{1.2in}|}{-} \\ \hline 
 &  & ${}^{233}$U & ${}^{233}$U & ${}^{239}$Pu \\ \hline 
Середня кількість народжених нейтронів в одному акті поділу & $\nu $ & 2.52 & 2.52 & 2.91 \\ \hline 
Переріз поділу & $\sigma_f$, барн & 260.68 & 260.6 & 477.0 \\ \hline 
Переріз захоплення & $\sigma_c$, барн & 52.97 & 52.97 & 286.1 \\ \hline 
\multirow{6}{2in}{Період напіврозпаду ядер- попередників залежно від групи запізнілих нейтронів} & T${}_{1}$, c & 55 & 55 & 54.28 \\ \cline{2-5}
 & T${}_{2}$, c & 20.57 & 20.57 & 23.04 \\ \cline{2-5}
 & T${}_{3}$, c & 5 & 5 & 5.6 \\ \cline{2-5}
 & T${}_{4}$, c & 2.13 & 2.13 & 2.13 \\ \cline{2-5}
 & T${}_{5}$, c & 0.62 & 0.62 & 0.62 \\  \cline{2-5}
 & T${}_{6}$, c & 0.28 & 0.28 & 0.26 \\ \hline 
\multirow{6}{2in}{Частка запізнілих нейтронів в залежності від їх групи} 
 & $\beta_1$, $10^{-3}$ & 0.224 & 0.224 & 0.072 \\ \cline{2-5}
 & $\beta_2$, $10^{-3}$ & 0.776 & 0.776 & 0.626 \\ \cline{2-5}
 & $\beta_3$, $10^{-3}$ & 0.654 & 0.654 & 0.444 \\ \cline{2-5}
 & $\beta_4$, $10^{-3}$ & 0.725 & 0.725 & 0.685 \\ \cline{2-5}
 & $\beta_5$, $10^{-3}$ & 0.134 & 0.134 & 0.18 \\ \cline{2-5}
 & $\beta_6$, $10^{-3}$ & 0.087 & 0.087 & 0.093 \\ \cline{2-5}
 & $\beta_?$, $10^{-3}$ & 2.6 & 2.6 & 2.1 \\  \hline 
Довжина реактора\textit{} & \textit{l, }см & 100 & 100 & 100 \\ \hline 
\end{tabular}
\end{center}
\label{tab01}
\end{table}

Відзначимо, що перерізи нейтронно-ядерних реакцій для нуклідів, наведених вище, задавалися їх усередненими значеннями по області енергій нейтронів (0.015$\div$10 еВ).

\begin{figure}[ht]
\noindent\centering{\includegraphics[width=12cm]{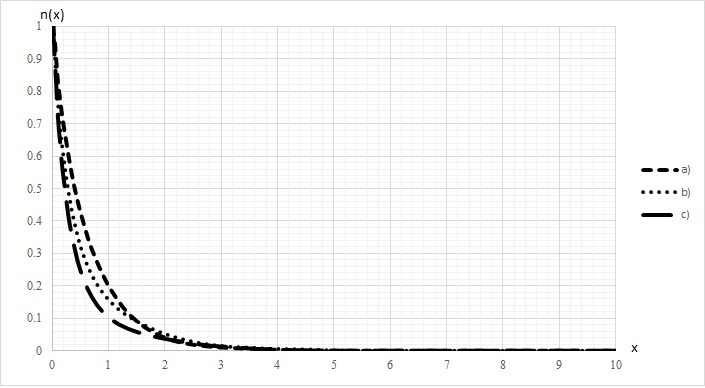}}
\vskip-3mm
\caption{Кінетика нейтронів при хвильовому нейтронно-ядерному горінні торію зі збагаченням по ${}^{239}$Pu, залежність знерозміреної щільності нейтронів від просторової координати $n(x)$ для моменту часу розрахунку:
а) \textit{t~=~}72~діб; b) \textit{t~=~}216~діб; c) \textit{t~=~}360~діб.}
\label{fig02}
\end{figure}

\begin{figure}[htbp!]
\center{\includegraphics[width=12cm]{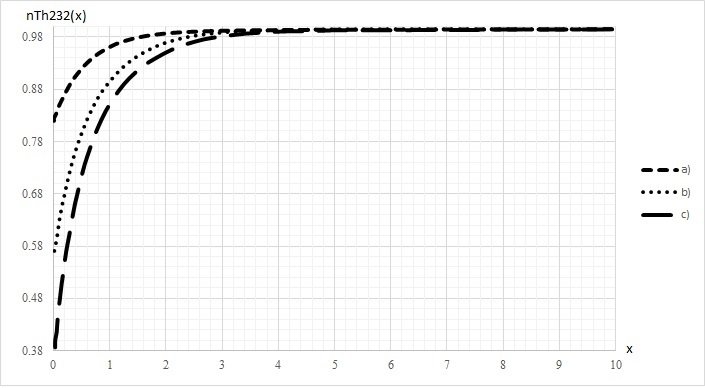}}
\vskip-3mm
\caption{Кінетика щільності ядер ${}^{232}$Th при хвильовому нейтронно-ядерному горінні торію зі збагаченням по ${}^{239}$Pu, залежність знерозміреної щільності ядер ${}^{232}$Th від просторової координати $N^{Th{\rm 232}} \; \left(x\right)$ для моменту часу розрахунку:
a) \textit{t =~}72~діб; b) з \textit{t =~}216~діб; c) \textit{t =~}360~діб.}
\label{fig03}
\end{figure}

\begin{figure}[htbp!]
\center{\includegraphics[width=12cm]{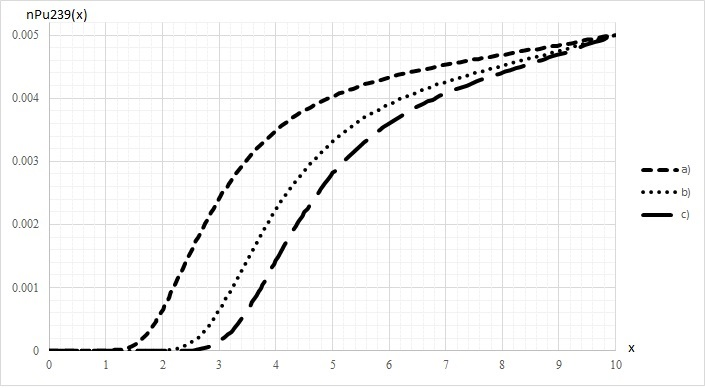}}
\vskip-3mm\caption{Кінетика щільності ядер ${}^{239}$Pu при хвильовому нейтронно-ядерному горінні торію зі збагаченням по ${}^{239}$Pu, залежність знерозміреної щільності ядер ${}^{239}$Pu від просторової координати $N^{{\rm Pu239}} \; \left(x\right)$ для моменту часу розрахунку:
a) \textit{t =~}72~діб; b) \textit{t =~}216~діб; c)  \textit{t =~}360~діб.}
\label{fig04}
\end{figure}

\begin{figure}[htbp!]
\center{\includegraphics[width=12cm]{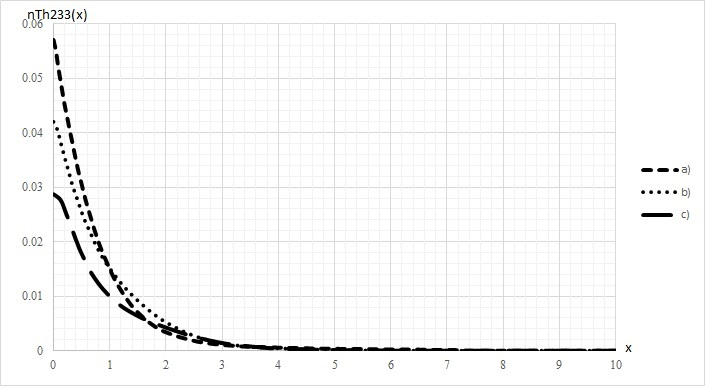}}
\vskip-3mm\caption{Кінетика щільності ядер ${}^{233}$Th при хвильовому нейтронно-ядерному горінні торію зі збагаченням по ${}^{239}$Pu, залежність знерозміреної щільності ядер ${}^{233}$Th від просторової координати $N^{Th{\rm 233}} \; \left(x\right)$ для моменту часу розрахунку:
a) \textit{t =}72 діб; b) \textit{t =~}216~діб; c)  \textit{t =~}360~діб.}
\label{fig05}
\end{figure}

\begin{figure}[htbp!]
\center{\includegraphics[width=12cm]{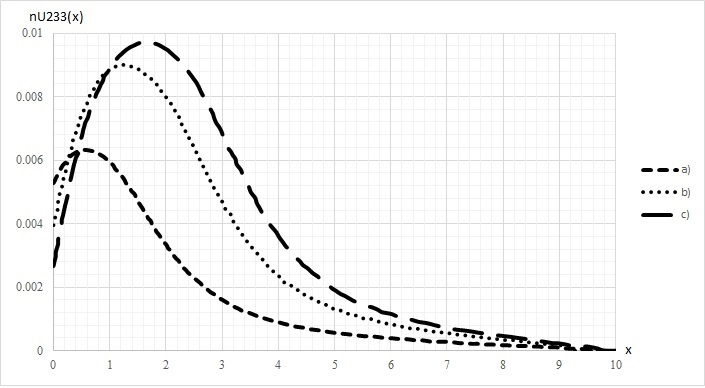}}
\vskip-3mm\caption{Кінетика щільності ядер ${}^{233}$U при хвильовому нейтронно-ядерному горінні торію зі збагаченням по ${}^{239}$Pu, залежність знерозміреної щільності ядер ${}^{233}$U від просторової координати $N^{U{\rm 233}} \; \left(x\right)$ для моменту часу розрахунку:
a)\textit{ t =~}72 діб; b)  \textit{t =~}216 діб; c)  \textit{t =~}360 діб.}
\label{fig06}
\end{figure}

При розрахунку, результати якого представлені нижче на рис.~\ref{fig02} - \ref{fig06}~a),~b),~c), використовуються значення із таблиці~\ref{tab01}, де час моделювання ХЯГ 360~діб, крок за часом $\Delta t=$50~хв, крок по просторовій координаті $\Delta x= 0.01$~см.

Звичайно, хотілося б провести розрахунок для значно більшого часу комп'ютерного експерименту, і щоб крок за часом мати $\Delta t\approx {\rm 10}^{-{\rm 5}} \div {\rm 10}^{-{\rm 7}} \; c$, але при виборі зазначених вище параметрів, що задаються при розрахунку, були обмежені наявними обчислювальними ресурсами.

Представлені на рис.~\ref{fig02}-\ref{fig06} результати чисельного моделювання хвильового нейтронно-ядерного горіння торію зі збагаченням по ${}^{239}$Pu в тепловій області енергій нейтронів (0.015$\div$10 еВ) свідчать про реалізацію такого режиму. Дійсно, на рис.~\ref{fig06} ми бачимо хвильове горіння ${}^{233}$U, що можна помітити по збільшенню концентрації нукліда, що поділяється, в зоні горіння, а згодом і зменшенням його концентрації при зростанні швидкості його вигоряння. А також на рис.~\ref{fig06} ми бачимо динаміку зміщення максимальної концентрації ${}^{233}$U по довжині палива у відповідності зі збільшенням часу горіння, тобто бачимо хвилю ядерного горіння, що біжить по паливу. При цьому згідно з рис.~\ref{fig03} і \ref{fig04}, ${}^{232}$Th і ${}^{235}$U поступово вигорають. 

Слід зазначити, що представлені на рис.~\ref{fig02} результати кінетики для щільності нейтронів не демонструють нейтронну хвилю. Пояснюється це тим, що чисельно вирішувалася система диференціальних рівнянь щодо знерозмірених (згідно співвідношенням (\ref{eq22})) змінних і при знерозмірюванні, щільність нейтронів ділилася на щільність потоку зовнішнього джерела, який при розрахунку задавався спеціально завищеним значенням з метою скоротити час розрахунку і рівним $\Phi_0 = 1.0 \cdot 10^{16}$ см${}^{-2}$с${}^{-1}$, тому відмінність масштабів величин щільності потоку зовнішнього джерела і щільності потоку нейтронів в області ядерного горіння в режимі усталеного хвильового горіння не дозволяє бачити нейтронну хвилю. 

Можливо також, що зміщення хвилі нейтронів уздовж просторової координати  не видно на рис.~\ref{fig02} при зазначеній щільності нейтронів, оскільки на нього накладається горіння ${}^{239}$Pu. Дійсно, результати моделювання кінетики щільності ядер ${}^{239}$Pu, представлені на рис.~\ref{fig04}, показують, що ${}^{239}$Pu бере участь в поділі для наповнення активної зони нейтронами на етапі підпалу, а його початкова концентрація дорівнює 0.5\% та лише через 72 доби вона стає менше на $\sim$0.15\%. Через 360 діб концентрація ${}^{233}$U стане вдвічі більшою (згідно з рис.~\ref{fig06} дорівнює \~{}1\%).

Підкреслимо те, що згідно з результатами, представленими на рис.~\ref{fig06}, хвиля повільного нейтронно-ядерного горіння ${}^{233}$U сформувалася за час моделювання, яке дорівнювало 360 діб.


\section{Висновки}
Представлені результати дослідження виконання критерію ХЯГ, тобто можливості хвильового ядерного горіння для середовища, що поділяється та спочатку складається з ${}^{232}$Th зі збагаченням 2\%, 1\% і 0.5\% по ${}^{239}$Pu, для області енергій нейтронів 0.015-10~еВ. Ці результати свідчать про виконання критерію ХЯГ, тобто про можливість реалізації режиму хвильового нейтронно-ядерного горіння торій-уранових паливних середовищ, що спочатку (до процесу ініціації хвильового режиму горіння за допомогою зовнішнього джерела нейтронів) мають збагачення по ${}^{239}$Pu, відповідні підкритичним станам в області теплових, надтеплових і проміжних нейтронів.

Вперше для підтвердження справедливості висновків з розділів \ref{sec01} і \ref{sec02} проведено моделювання хвильового ядерного горіння вказаного вище паливного торієвого середовища протягом 360 діб для діапазону енергій нейтронів (0.01-10~еВ). Отримані та представлені у роботі результати доводять реалізацію режиму хвильового ядерного горіння в області теплових та надтеплових енергій.

Перевагами такого паливного середовища з торію для реактора, який буде працювати у режимі хвилі ядерних поділів, що біжить, є: 
\begin{itemize}
\item саморегулювання хвилі за часовим параметром, який дорівнює майже 40 діб, що значно перевищує такий само часовий параметр для ХЯГ в уран-плутонієвому паливному середовищі, який дорівнює 3.3 добам, що одночасно вказує на більшу безпеку такого режиму горіння; 

\item при поділі торію утворюється значно менша кількість довговічних продуктів поділу та актинідів, таких як нептуній, америцій і самарій. Вихід нептунію, америцію і самарію в $\sim 10^2$, $\sim 10^5$ і $\sim 10^{6}$ разів відповідно менше ніж при урановому циклі. Так само після 10-річного відстою використаного палива розпадається більшість продуктів поділу ${}^{233}$U \cite{19}, що значно спрощує завдання локалізації та захоронення РАВ \cite{19}. 

\item приблизна концентрація торію \cite{19,20} в земній корі в 3-4 рази більша, ніж урану.

\item використання палива із торієвого сольового розтопу дасть можливість повністю уникнути використання легкої води в якості теплоносія, яка може розкластися на кисень і вибухонебезпечний водень.
\end{itemize}

Найбільш цікавими є області енергій нейтронів від 0.015-10~еВ, оскільки при даних значеннях знімається питання радіаційної стійкості матеріалів \cite{3,15} (оболонки активної зони) і відповідно, меншої товщини захисної оболонки.

Однак існують проблеми \cite{19} при використанні паливної матриці у вигляді металу або двоокису торію. Ц пов'язано з протактинієм-233, який є добрим поглиначем нейтронів (час життя близько 27 діб). Оскільки цей нуклід знаходиться в ланцюжку розпаду при напрацюванні ${}^{233}$U (\ref{eq01}), то при поглинанні нейтрона ${}^{233}$Pa ми будемо втрачати цінні для ХЯГ ядра ${}^{233}$U. Відповідно, даний проміжний нуклід на шляху напрацювання подільного нукліда потрібно виводити з активної зони на час періоду його напіврозпаду. Даний процес неможливо здійснити, якщо паливна матриця буде мати тверду структуру, і відповідно, кращим буде паливне середовище із сольовоого розтопу, для якого вже існує технологічний процес з виведення його з активної зони \cite{20}.

\abstract
\begin{center}
\title{Simulation of the wave neutron-nuclear burning of Th232 enriched with Pu239 for the thermal range of neutron energy}
\end{center}

\begin{center}	
\author{A.O.Kakaev,  V.O. Tarasov, S.A. Chernezhenko, V.D. Rusov$^{\mathbf{1}}$}
\author{ V. O. Sova$^{\mathbf{2}}$}
		
\textit{Odessa National Polytechnic University, Shevchenko Ave. 1, Odessa, 65044, Ukraine, e-mail:andreykakaev@gmail.com}\\
\textit{State Scientific and Technical Center for Nuclear and Radiation Safety, Vasyl-Stus str. 35-37, 03142, Kiev, Ukraine}\\

	\end{center}

At the initial stage, the development of wave reactors, which will operate in the mode of wave nuclear burning (WNC), requires the study of the kinetics of the WNC fuel mode when changing both external parameters (flux density of an external neutron source, thermal parameters of heat transfer) and internal parameters (fuel composition, reactor material parameter, delayed neutrons).
In order to confirm the possibility of WNC with enrichment, we study its impact on the criterion of WNC for thorium fuel (${}^{232}$Th at different enrichments in ${}^{239}$Pu) and its behavior during the ``ignition'' stage.
To confirm the fulfillment of the criterion of slow WNC depending on the neutron energy, we perform a numerical simulation of the thorium fuel WNC mode dynamics, taking into account the delayed neutrons in the thermal and epithermal regions of the neutron energies (0.015-10~eV).
Keywords: thorium reactor fuel, criterion of wave nuclear burning, simulation of wave nuclear burning, reactor that operates in the mode of wave nuclear burning.

\end{document}